\documentclass[12pt]{article}
\usepackage{graphicx}
\usepackage{amssymb}

\def\Title#1{\begin{center} {\Large #1 } \end{center}}
\def\Author#1{\begin{center}{ \sc #1} \end{center}}
\def\Address#1{\begin{center}{ \it #1} \end{center}}

\begin{document}

\Title{\textbf{C-parity and Regge Intercepts\footnote{The early, very concise and incomplete version of this work can be found in \cite{Ptr} } .}}

\Author{Vladimir A. Petrov\footnote {E-mail: {Vladimir.Petrov@ihep.ru}}  } 
\Address{ Logunov Institute for High Energy Physics, \\ NRC Kurchatov Institute, \\ \textbf{1} Nauki square, Protvino, 142281,Russian Federation}

 \begin{abstract}
In this work, we obtain constraints on the intercepts of the leading Regge pole trajectories - Pomeron and Odderon- implies by C-parity and based on the heuristic argument that C-even exchanges correspond to attractive forces between hadrons undergoing elastic scattering, and C-odd exchanges correspond to attractive forces in the case of particle-antiparticle scattering and repulsive forces in the case of particle-particle scattering. 

In addition, the case of secondary degenerate trajectories is also considered and a lower bound for their intercept is obtained.

\end{abstract}

\begin{center}
\section{Introduction}
\end{center}

Motivation and impetus for conducting this study is a long-standing debate regarding the properties of the C-odd partner of the Pomeron, the Odderon. 
Since the interaction (quasi)potential\footnote{As a quasi-potential in the quasipotential equation \cite{Log}  as a relativistic generalization of single-particle exchange, we take the single-Reggeon exchange in the spirit of L.Van Hove's construction \cite{Van}.}, by which one can judge the nature of inter-nucleon forces, is usually associated with the eikonal phase, in what follows we will consider the eikonal phase as being determined at high energies by Born amplitudes with single-Reggeon exchanges.
For the sake of completeness, we have also included the Pomeron, Odderon's "older half-brother", in our consideration.
In the following we will limit ourselves to unpolarized  scattering.

\section{Born amplitudes}

 As stated above, we define the Pomeron and the Odderon as Regge pole-trajectories determining the high-energy behaviour of the Born amplitudes of scattering processes. Note that, unlike secondary Regge trajectories ($ \rho,\omega $, etc.), there are currently no states ("glueballs") observed to date that could be considered to lie on these two "vacuum" trajectories\footnote{In principle, trajectories that do not intersect integer values are conceivable, but we leave this option aside for now.} , although some mesons have been discussed in this regard. Unfortunately, despite enormous efforts over several decades, not only have Regge trajectories not been derived as definite functions within the framework of QCD, but there are not even any reliable theoretically justified values for intercepts and slopes at $ t=0 $. 
 
 The total scattering amplitudes as functions of complex angular momenta $ T(j,t) $ have as leading singularities not poles, but branching points at $j= j_{c}(t)$. Note that $ s $-channel unitarity (more precisely the Froissart-Martin upper bound) implies that $ Re j_{c}(t)\leq 1 $ for physical transfers  $ t\leq0 $.
Meanwhile, there is no upper limit for the intercepts of Regge-pole trajectories $ \alpha(t=0)\equiv 1+\Delta $ so 
$ \Delta $ can be both positive and negative\footnote{In the early days of Reggeism, it was believed that the Pomeron intercept could not exceed 1, but the eikonal approach as a way to sum up many-Pomeron exchanges removes this limitation.}.

After the discovery of the increase of total cross sections (first in $ K^{+}p $  \cite{Den} and then in $ pp $, \cite{Mnd} ,  \cite{Am}) the $ \Delta_{\mathcal{P}} > 0 $ option prevails for the Pomeron (C-even vacuum exchange), although in some models (for example, "Reggeon field theory") the option $ \Delta_{\mathcal{P}} =0 $ is also retained, allowing, nonetheless, for an unlimited growth of cross sections \cite{Ter}, albeit in a very remote asymptotic range.

As for the C-odd partner of the Pomeron, the notorious Odderon, which we will also consider a Regge-pole trajectory, all possible options were also presented for its intercept $ \alpha_{\mathcal{O}}(0)\equiv 1+\Delta_{\mathcal{O}} $, i.e. $ \Delta_{\mathcal{O}}> 0,\Delta_{\mathcal{O}}< 0 ,\Delta_{\mathcal{O}}=0 $ \cite{Ew},\cite{Ew'}. 
Recent refinements concerning Odderon along 
the line of the rft \cite{Ter} can be found in \cite{Bra}. 
A mini review on the Odderon see in \cite{Ry}.

The standard expressions for the Born amplitudes\footnote{to use their Fourier-Bessel images as eikonal phases in full amplitudes in the eikonal scheme (see the next Section).}  (as already said above, further we will consider $ \bar{p}p $ and $ pp $ scattering) with Pomeron and Odderon exchanges are \cite{Van} 
\begin{equation}
\hat{\delta}  ^{\bar{p}p,pp}_{\mathcal{P}}(s,t)=-g_{\mathcal{P}}^{2}(t)\frac{1+exp(-i\pi\alpha_{\mathcal{P}}(t))}{sin\pi\alpha_{\mathcal{P}(t)} }s^{\alpha_{\mathcal{P}}(t)}
\end{equation}

\begin{equation}
\hat{\delta}^{\bar{p}p}_{\mathcal{O}}(s,t)= g_{\mathcal{O}}^{2}(t)\frac{1-exp(-i\pi\alpha_{\mathcal{O}}(t))}{sin\pi\alpha_{\mathcal{O}(t)} }s^{\alpha_{\mathcal{O}}(t)}
\end{equation}

\begin{equation}
\hat{\delta}^{pp}_{\mathcal{O}}(s,t)= - g_{\mathcal{O}}^{2}(t)\frac{1-exp(-i\pi\alpha_{\mathcal{O}}(t))}{sin\pi\alpha_{\mathcal{O}(t)} }s^{\alpha_{\mathcal{O}}(t)}
\end{equation}

\begin{center}
\section{Potentials}
\end{center}

Potentials are defined as follows \footnote{Upper dots mean $ \bar{p}p $ or $ pp $ while the lower ones $ \mathcal{P} $ or $\mathcal{O}$. } 

\begin{equation}
V^{...}_{...}(s,r)= -\frac{1}{2s}\int \frac{d^{3}q}{(2\pi)^{3}} e^{irq} \hat{\delta}^{...}_{...}(s,t=-\textbf{q}^{2})
\end{equation}
They are related to the eikonal phases $ \delta(s,b) $ as follows
\begin{equation}
\delta(s,b)= - \frac{1}{2} \int dz V^{...}_{...}(s,r=(z,\textbf{b}))= \frac{1}{8\pi s}\int dq q J_{0}(bq)\hat{\delta}^{...}_{...}(s,t=-\textbf{q}^{2})
\end{equation}
Let us recall that the total scattering amplitude $ T^{{\bar{p}p,pp}}(s,t) $ is related to \\

\begin{center}
$ \delta^{...}(s,b) = \sum_{k=\mathcal{P},\mathcal{O},... }\delta_{k}^{...}(s,b)  $
\end{center}
 as follows:
\[T^{...}(s,t)=4i\pi s\int_{0}^{\infty}db b J_{0}(b\sqrt{-t})\tilde{T}(s,b)  \equiv 4i\pi s\int_{0}^{\infty}db b J_{0}(b\sqrt{-t})(1-e^{2i\delta^{...}(s,b)})\]

From the unitarity condition
\begin{equation}
\Im \tilde{T}(s,b)\geq \mid {\tilde{T}}(s,b)\mid^{2}
\end{equation}
 it follows that $ \Im (1-e^{2i\delta^{...}(s,b)})\geq 0 $
and hence 
\begin{equation}
\Im \delta^{...}(s,b)\geq 0.
\end{equation}
In the Regge regime (high energies, low momentum transfers) the approximate expressions for (1)-(3), that we will use, are:

\begin{equation}
\hat{\delta}^{\bar{p}p,pp}_{\mathcal{P}}(s,t)\approx g_{\mathcal{P}}^{2}[i + \tan (\frac{\pi\Delta_{\mathcal{P}}}{2}2)]s^{1+\Delta_{\mathcal{P}}}exp(R^{2}_{\mathcal{P}}t/4)
\end{equation}

\begin{equation}
\hat{\delta}^{\bar{p}p}_{\mathcal{O}}(s,t)\approx g_{\mathcal{O}}^{2}[i - \cot \frac{\pi\Delta_{\mathcal{O}}}{2}]s^{1+\Delta_{\mathcal{O}}}exp(R^{2}_{\mathcal{O}}t/4)
\end{equation}

\begin{equation}
\hat{\delta}^{pp}_{\mathcal{O}}(s,t)\approx - g_{\mathcal{O}}^{2}[i - \cot \frac{\pi\Delta_{\mathcal{O}}}{2}]s^{1+\Delta_{\mathcal{O}}}exp(R^{2}_{\mathcal{O}}t/4)
\end{equation}
where 
\begin{equation}
R^{2}_{\mathcal{P,O}}=4\alpha'_{P,O}\ln s + 2r^{2}_{P,O},
\;r^{2}_{P,O}=4\partial (ln(g_{\mathcal{P,O}}^{2}(t))/\partial t \mid_{t=0}.
\end{equation}
The apparent singularity at $ \Delta_{\mathcal{O}}=0 $  in Eqs.(9)$ \div $(10) is avoided as discussed in detail below Eq.(21)\footnote{ A detailed discussion of this special case see in \cite{Pit}. } .

From the impact images of the eikonals we get 

\begin{equation} 
\Im \delta^{pp}(s,b)=g_{\mathcal{P}}^{2}\frac{s^{\Delta_{\mathcal{P}}}exp(-b^{2}/R^{2}_{\mathcal{P}})}{4\pi R^{2}_{\mathcal{P}}}-g_{\mathcal{O}}^{2}\frac{s^{\Delta_{\mathcal{O}}}exp(-b^{2}/R^{2}_{\mathcal{O}})}{4\pi R^{2}_{\mathcal{O}}}.
\end{equation}
Inequality (7) is satisfied only when
\[\Delta_{\mathcal{P}}>\Delta_{\mathcal{O}}. \]
Now we can move on to potentials.\\The Pomeron one is
\begin{equation}
V_{\mathcal{P}}^{\bar{p}p,pp}\approx -g^{2}_{P}(i+tan\frac{\pi\Delta_{\mathcal{P}}}{2})V_{\mathcal{P}}(r)
\end{equation}
while the Odderon ones are dependent on initial state
\begin{equation}
V^{\bar{p}p}_{\mathcal{O}}\approx -g^{2}_{O}(i-cot\frac{\pi\Delta_{\mathcal{O}}}{2})V_{\mathcal{O}}(r),
\end{equation}

\begin{equation}
V^{pp}_{\mathcal{O}}\approx g^{2}_{O}(i-cot\frac{\pi\Delta_{\mathcal{O}}}{2})V_{\mathcal{O}}(r).
\end{equation}
where $V_{\mathcal{P}}(r)$ and $ V_{\mathcal{O}}(r) $ are positive decreasing functions of $ r $ the detailed dependence of which on $ r $ is determined by Regge factors $ s^{\alpha_{...}(t)} $ and vertex functions $ g_{...}(t) $.

At very large $ r $ 
\[V_{\mathcal{P}}(r)\sim exp(-2m_{\pi}r),V_{\mathcal{O}}(r)\sim exp(-3m_{\pi}r). \]

From Eq.(9) we see that attractive force can be formally realized at 

\begin{center}
$ 2n < \Delta_{\mathcal{P}}< 2n+1, n\in\mathbb{Z} $
\end{center}
where $ \mathbb{Z} $ is the set of all integers, including negative ones. There are no solid theoretical grounds for choosing from this diversity, and we, for lack of anything better, will be guided by phenomenology, which tells us that
\begin{equation}
0 < \Delta_{\mathcal{P}}< 1.
\end{equation}

In work \cite{Lip} a boundary value was obtained
\begin{equation}
\alpha_{\mathcal{P}}(t)\rightarrow 1, -t\rightarrow\infty.
\end{equation} 
which is in complete agreement with the picture of the exchange of non-interacting gluons at short distances. 
This is also supported by the Gribov-Pomeranchuk formula \cite{Squi} (see also \cite{Col})   for the position of the singularity due to exchange of arbitrary number of identical Reggeons:
\begin{equation}
\alpha_{n} = n\alpha(t/n^{2})-n+1.
\end{equation}
where $ n $ is any integer.
In case of non-interacting gluons $ \alpha(t) = 1 $ so

\begin{center}
$ \alpha_{n} =1,  \forall n. $
\end{center}

Assuming that 
\begin{center}
$ \Im \alpha_{\mathcal{P}}\geq 0 $ 
\end{center}
(as is the case in all specific examples and obviously in potential scattering)
we obtain from the dispersion relation
\begin{equation}
\alpha_{\mathcal{P}}(t)= 1 + \frac{1}{\pi}\int_{4m_{\pi}^{2}}^{\infty}dt'\Im \alpha_{\mathcal{P}}(t')/t'-t)
\end{equation}
that the negative option for $\Delta_{\mathcal{P}} $ is unacceptable without violation of Eq.(17). Of course, from the point of view of phenomenology this is obvious. For the same reason, option  $ \Delta_{\mathcal{P}}> 2 $ also seems unrealistic.
Thus, we remain with the "supercritical" Pomeron with
$ 0 <\Delta_{\mathcal{P}}< 1 $.
Actually, phenomenologically used $ \Delta_{\mathcal{P}} $'s are located closer to the left edge

\begin{center}
$ \Delta_{\mathcal{P}}\approx 0.07\div 0.30 .$
\end{center}
Now let us turn to the Odderon.\\From Eq.(15) we conclude that the Odderon gives attraction between $ p $ and $ \bar{p} $ at cases 

\begin{equation}
2n-1< \Delta_{\mathcal{O}} < 2n, n \in \mathbb{Z}.
\end{equation}
It follows from the unitarity condition that   $ \Delta_{\mathcal{O}} \leq \Delta_{\mathcal{P}} $.

Again taking into account the phenomenology, we come to the conclusion that

\begin{equation}
-1< \Delta_{\mathcal{O}} \leq 0 .
\end{equation}
As expected, in this case there is repulsion between $ p $ and $ p $ if $\Delta_{\mathcal{O}} < 0  $.

The case $ \Delta_{\mathcal{O}}= 0 $ is special and requires a separate consideration. Now, Eq. (7) modifies to

\begin{equation}
\hat{\delta}^{\bar{p}p}_{\mathcal{O}}(s,t)\approx g_{\mathcal{O}}^{2}[i - \frac{2}{\pi\alpha^{'}_\mathcal{O}t}]s^{1+{\alpha^{'}_\mathcal{O}}t}exp(r^{2}_{\mathcal{O}}t/2)
\end{equation}
Eq.(22) reveals a strong interacting neutral massless vector ("strong photon") in the cross-channel.
To avoid such a bizarre prediction we have to ask for a zero in the vertex $ g_{\mathcal{O}}(t) $ at $ t=0 $. From some analyticity considerations\footnote{The factor $ g_{\mathcal{O}}  $ stems from the vertex $ \Gamma (J=1,t)$ describing the nucleon-Odderon coupling is analytic at $ t=0 $. Details are presented in \cite{Pit} .}  this zero must be of integer order not less than 1
\begin{equation}
g_{\mathcal{O}}(t)\mid_{t\rightarrow 0} \approx t^{N}\gamma(0), \gamma(0)\neq0, N\geq1.
\end{equation}

Thus, we see that the option $ \alpha_{\mathcal{O}}(0)= 1 $ leads, after killing of a "strong photon" in the cross-channel, to decoupling of the Odderon from the forward scattering at the level of Born amplitudes. In the full scattering amplitude $ T_{\mathcal{P}+\mathcal{O}} (s,t)$ the Odderon will appear only as in a convolution with the Pomeron, which is not literally zero at $ t=0 $ but reduces the chances of observing the Odderon effect in the forward region.

To complete this Section, let us note that comparison of the imaginary parts of the complex potentials, with account of obtained properties of the Pomeron and Odderon intercepts, shows that central absorption is stronger in the $ \bar{p}p $ channel than in the $ pp $ channel:
\begin{equation}
-\Im V_{\bar{p}p}\sim g^{2}_{P} + g^{2}_{O},
\end{equation}
\begin{equation}
-\Im V_{pp} \sim  g^{2}_{P} - g^{2}_{O}.
\end{equation} 
This is dictated by the consequence of the s-channel unitarity:
\[\Im V \leq 0 \rightarrow g^{2}_{P} \geq g^{2}_{O}.\]

\section{Unexpected byproduct or what is the reason for the degeneration of secondary trajectories?}

Above, we dealt with "vacuum" trajectories: Pomeron and Odderon.
However, it turns out that interesting results can also be obtained for "secondary" trajectories.
In the literature, it is generally accepted that $ \omega $ and $ f $ trajectories are degenerate because there should be no "exotic" (in the terminology of dual models of the beginning of 1970s) channels. At the same time, the caveat was always added: “at high energies.”
In fact, from dual diagrams in the spirit of Harari \cite{Ha} - Rosner \cite{Ros}   , it is clear that, for example, in $ pp $ or, say, $ K^{+}p $ scattering or else, only Pomeron and Odderon contribute to the imaginary part of the amplitude; cuts with a net colour would lead to confinement violation, independently of the channel energy. 

This immediately implies the degeneracy of $ f $ and $ \omega $ trajectories as well as trajectories $ A_{2} $ and $ \rho $, as well as $ \varphi $ and $ f^{'} $ etc.

Now, a problem seems to arise: if the trajectory is neither C-even nor C-odd, then how do we determine what the nature of the forces between hadrons, due to such exchange, will be?

To get rid of the uncertainty, we assume that the parity of the forces should be determined by the C-parity of the lowest excitation on the degenerate trajectory.
Accordingly to this reasoning we get

\begin{equation}
\hat{\delta}^{\bar{p}p}_{R}(s,t)= -g^{2}_{R} (-i + \cot \pi \alpha_{R}(t))s^{\alpha_{R}(t)};\\\hat{\delta}^{\bar{p}p}_{R}(s,t)= -g^{2}_{R} (\csc (\pi \alpha_{R}(t))s^{\alpha_{R}(t)}
\end{equation}
Note the pure real contribution in $ pp $ case.
Reasoning similar to that given in the previous section leads to the following inequalities:
\begin{equation}
0.5 \leq \alpha_{R}(0)\leq 1, R = \omega,\rho^{0},\varphi... 
\end{equation}
These limits, quite trivial from the phenomenology view,  mean,however, that the corresponding exchange lead to repulsion in $ pp, K^{+}p,... $ channels and to attraction in the $ \bar{p}p,K^{-}p,... $-channels.

Let us note that all the considerations above concern trajectories containing light-quark quarkonia.
In the case of mesons $ H $ with heavy quarks one 
should choose 
\begin{equation}
 \alpha_{H}(0)< 0
\end{equation}
The specific values of intercepts in this case require special consideration \cite{Ser} (further development see in \cite{Ger})
and will be dealt elsewhere. 
\section{Conclusions and discussion}

In this paper, relying on the requirement of attractive C-even forces and repulsive C-odd forces corresponding to the leading Reggeons, Pomeron and Odderon, we concluded that:

a) the Pomeron intercept is of the form $ 1+\Delta_{\mathcal{P}}, \Delta_{\mathcal{P}}> 0 $;

b) the Oddderon intercept cannot exceed unity but most\\ 
likely goes higher than $ -1 $.

Despite the arguments presented above, the negative value of 
$ \Delta_{\mathcal{O}} $  causes the author some concern because, unlike the Pomeron, we do not yet know how the Odderon trajectory behaves in the deep Euclidean region. Naive considerations (exchange by three noniteracting gluons) suggest that
\begin{center}
$ \alpha_{\mathcal{O}} (t)\rightarrow 1 , -t\rightarrow \infty.$
\end{center}
However, from the dispersion relations for the Odderon trajectory of type (19) with non-negative $ \Im \Delta_{\mathcal{O}}(t) $ we see that in this case $ \Delta_{\mathcal{O}} $ should be \textit{positive} in contradiction with Eq.(19).
The contradiction can be resolved either by assuming that $ \Im \Delta_{\mathcal{O}}(t) $ is not non-negative definite or by assuming that the deep Euclidean limit for the Odderon lies below 1.
Currently, there is still no answer to either of these statements.

The studies of the Odderon with a maximum intercept at 1 ($ \Delta_{\mathcal{O}}=0 $)  were made in Ref.\cite{Vac} in the framework of the notorious scheme "BFKL", although the corresponding singularity in the $ j $-plane (the Born level) is apparently not a pole, but a root cut.

The option with a fixed ($ t $-independent) "critical" Regge-pole Odderon was considered  in the work \cite{Lu}. Although, of course, such a singularity is hardly acceptable.

"Supercritical" Odderon with $ \Delta_{\mathcal{O}}> 0 $ lies in the base of so-called "Maximal Odderon" model \cite{Nic}. This approach contradicts our reasoning while the physical content and descriptional quality of the very model was criticized in Ref.\cite{Pet}.

A "byproduct" of our reasoning , Eq.(27), seems to reject a long-noticed tendency for the $f $-trajectory (in this case  not degenerated with $ \omega $ trajectory) to have a greater intercept than other trajectories\footnote{At one time, there was even a lively debate about the identity of the Pomeron and the $ f $-trajectory \cite{Ch}, \cite {Chu}.}. Such an option would contradict to our argument based on confinement, so we cannot support it. 

It should be noted that for phenomenological purposes   parameters with $ \alpha_{R}(0) < 0.5 $ often used as a result of  keeping linearity of the trajectories beyond the domain of  quasi-classicality.

Our lower bound $ \alpha_{R}(0) > 0.5 $ values is pretty possible and my be well related with the non-linearity of the trajectories (see for example \cite{Ji})in the region near $ t=0 $, which is not a region of quasi-classicality leading to a linear dependence of the spin and the square of the mass.

We also note once again that, despite great experimental \cite{Ro}   and theoretical efforts, the \textit{quantitative} characteristics of the Pomeron and Odderon trajectories remain a subject of further research and discussion, to which this work is a contribution (we hope not without benefit).

\section*{Acknowledgements}

I thank the participants of the seminar of the Department of Theoretical Physics of the Institute for an interesting and useful discussion and the referees for valuable remarks.

\end{document}